\pdfoutput=1
\documentclass[floatfix,a4paper,aip,jcp,reprint]{revtex4-1}  
\usepackage{cmap}
\usepackage[T1]{fontenc}

\usepackage{mathptmx}
\usepackage{times}
\usepackage{fixmath}
\usepackage[Euler]{upgreek}
\DeclareSymbolFont{UPM}{U}{eur}{m}{n}  
\DeclareMathSymbol{\partial}{0}{UPM}{"40}
\usepackage[italic]{mathastext}
\usepackage{helvet}

\usepackage{mleftright}

\usepackage{booktabs}

\usepackage[pdftex,bookmarks, pdffitwindow=false, pdfstartview=FitH, pdfdisplaydoctitle, colorlinks, plainpages=false,pdftitle={Self-assembly of two-dimensional binary quasicrystals: A possible route to a DNA quasicrystal},pdfauthor={Aleks Reinhardt, John S. Schreck, Flavio Romano, Jonathan P. K. Doye}, pdfpagelabels, hypertexnames, citecolor={blue},linkcolor={red}, urlcolor={violet}, pdflang={en}, hyperfootnotes=false, breaklinks]{hyperref}

\usepackage{graphicx}
\usepackage{amsmath}

\usepackage{geometry}
\usepackage[dvipsnames]{xcolor}

\usepackage{setspace}
\usepackage[stretch=15,shrink=15,step=3]{microtype}
\usepackage[version=3,arrows=pgf]{mhchem}
\usepackage{siunitx}

\usepackage{flafter}

\newcommand{\refSub}[2]{\hyperref[#2]{\ref{#2}(#1)}}

\begin{document}

\title{Self-assembly of two-dimensional binary quasicrystals: A possible route to a DNA quasicrystal}
\author{Aleks Reinhardt}
\affiliation{Department of Chemistry, University of Cambridge, Lensfield Road, Cambridge, CB2 1EW, United Kingdom}
\author{John S.~Schreck}
\author{Flavio Romano}
\altaffiliation{Current address: Dipartimento di Scienze Molecolari e Nanosistemi, Universit\`{a} Ca' Foscari,  Via Torino 155, 30172 Venezia Mestre, Italy.}
\author{Jonathan P.~K.~Doye}
\affiliation{Physical and Theoretical Chemistry Laboratory, Department of Chemistry, University of Oxford, South Parks Road, Oxford, OX1 3QZ, United Kingdom}

\begin{abstract}
We use Monte Carlo simulations and free-energy techniques to show that binary solutions of penta- and hexavalent two-dimensional patchy particles can form thermodynamically stable quasicrystals even at very narrow patch widths, provided their patch interactions are chosen in an appropriate way. Such patchy particles can be thought of as a  coarse-grained representation of DNA multi-arm `star' motifs, which can be chosen to bond with one another very specifically by tuning the DNA sequences of the protruding arms. We explore several possible design strategies and conclude that DNA star tiles that are designed to interact with one another in a specific but not overly constrained way could potentially be used to construct soft quasicrystals in experiment. We verify that such star tiles can form stable dodecagonal motifs using oxDNA, a realistic coarse-grained model of DNA.
\end{abstract}




\maketitle 

\section{Introduction}
Since their discovery was reported in 1984,\cite{Shechtman1984} quasicrystals have been extensively studied, and many have unusual electronic\cite{Poon1992} and surface\cite{Rivier1993} properties. While most quasicrystals reported thus far have been metallic alloys,\cite{Steurer2004, Steurer2012} such structures have also increasingly been seen in soft matter systems,\cite{Zeng2005, Tsai2003, Dotera2011} for example in colloidal\cite{Fischer2011}  and micellar\cite{Xiao2012,Zeng2004} systems and polymer melts.\cite{Hayashida2007, Zhang2012,Gillard2016} Moreover, quasicrystals have been observed in a number of computer simulations.\cite{Widom1987, Leung1989, Dzugutov1993, Skibinsky1999, Engel2007, Keys2007, Johnston2010b, Johnston2010c, ZhiWei2012, Dotera2012,  HajiAkbari2009, HajiAkbari2011, HajiAkbari2011b, Iacovella2011, Kiselev2012, VanDerLinden2012, Reinhardt2013, Dotera2014, Barkan2014, Engel2015, Pattabhiraman2015, Jiang2015, Archer2015} A mean-field approach has shown that dodecagonal quasicrystals are often thermodynamically more stable than other types of quasicrystal,\cite{Narasimhan1988} and a theoretical approach suggests that quasicrystals in soft matter are likely to be dodecagonal.\cite{Lifshitz2007}

We have recently observed an example of such a dodecagonal soft quasicrystal when studying the self-assembly behaviour of a two-dimensional patchy-particle system.\cite{VanDerLinden2012} We studied a system of particles with an angular dependence, such that each particle had five attractive `arms', or patches, which could bond with other particles. For particles with fairly narrow patches, the system forms a crystal where each particle has a co-ordination number of five, albeit in an arrangement that must deviate from perfect five-fold symmetry. For sufficiently wide patches, a competition is set up between such a non-uniform pentavalent co-ordination and the hexagonal co-ordination characteristic of crystals of spherically symmetric particles. These hexa- and pentavalent environments form square--triangle tilings, which are known to be capable of forming dodecagonal quasicrystals.\cite{Gaehler1988,Oxborrow1993} Indeed, we observed precisely such quasicrystals in brute-force simulations.\cite{VanDerLinden2012} In order to confirm  whether such quasicrystals are stable rather than just kinetic products, we have also computed explicit phase diagrams for this one-component patchy-particle system.\cite{Reinhardt2013}  To do this, we used Frenkel--Ladd integration\cite{Frenkel1984} and direct-coexistence simulations\cite{Vega2008} to compute the free energies of the quasicrystal and its competing phases: the quasicrystal was found to be a robust feature of the system and it persisted as the thermodynamically stable phase over a range of parameterisations of the model and occupied significant portions of the phase diagrams we computed.

In the light of these results, we anticipated that it might not be overly difficult to self-assemble such quasicrystalline structures in experiment. Whilst a true `patchy particle' system\cite{DeVries2007, Cho2007, Yang2008, Kraft2009,  Wang2008, Mao2010, Duguet2011, Wang2012, Chen2011, Glotzer2007,Pawar2010, Liu2016}  confined in two dimensions, perhaps by density mismatching,\cite{Chen2011} would perhaps be the most obvious candidate, one attractive alternative might be to make use of DNA multi-arm motifs,\cite{Yan2003, He2005b, He2006, Zhang2008, Zhang2013, Zhang2016} which have been shown to be able to self-assemble into a range of effectively two-dimensional structures. DNA multi-arm motifs, or `star tiles', are DNA structures which form a star shape with a certain number of protrusions called `arms'. The DNA strands in these structures are complementary such that the bulk of the structure is fully bonded, but the very ends of each of the arms contain unpaired strands which can bond with other star tiles. Since the bonding between the tiles is mediated by DNA, one important advantage of this approach is that the bonding can be chosen to be as generic or as specific as we wish, simply by selecting appropriate DNA sequences.  A similar approach involves the construction of multi-arm motifs using DNA origami.\cite{Zhang2015,Wang2016} Using DNA tiles to construct quasicrystals would be broadly similar to the recently observed lanthanide-directed self-assembly of quasicrystals,\cite{Ecija2013,*Urgel2016} but the underlying framework is different in the sense that DNA star tiles are effectively `patchy particles' with varying numbers of arms, whereas the basic units in the lanthanide-directed self-assembly approach are point vertices (the metal) and separate edges (molecular linkers).

The simple patchy-particle model we have previously introduced describes much of the fundamental physical behaviour of the DNA star tiles; however, unlike colloidal patchy particles with wide patches, DNA star tiles have a well-defined valence, determined by the number of arms, and so a five-arm star tile cannot bond with six neighbours.
Therefore, the DNA tiles best map onto patchy particles with a narrow patch width, and there is no parameter equivalent to the patch width that could be varied in order to facilitate five-arm DNA tiles to form quasicrystals. Indeed, experimentally, five-arm DNA star tiles have been observed just to form the same two-dimensional crystalline arrays with pentavalent co-ordination as the five-patch particles do at narrow patch width.\cite{Zhang2008}

In this work, we suggest a potential means to enable DNA star tiles to self-assemble into a variety of structures at low temperatures, including a quasicrystalline phase. Rather than rely on a competition between hexavalent and pentavalent environments corresponding to patchy particles with five patches, as we have done in previous work,\cite{Reinhardt2013} here, we simulate a two-component mixture of patchy particles with five and six patches of the appropriate composition. We show that such mixtures continue to exhibit stable quasicrystals.

Although the behaviour of patchy particles maps onto star tiles perhaps surprisingly well,\cite{Doye2007,Wilber2009b} this level of abstraction may seem rather extreme. Furthermore, given that quasicrystals to the best of our knowledge do not appear to have been observed with DNA star tiles in experimental work,\footnote{A scaffolded  approach has been used to produce a `quasicrystalline' patch of a finite size;\cite{Zhang2015} however, such a finite structure by construction cannot grow and is not a phase in the thermodynamic sense.} it is important to investigate whether the simple coarse-grained model we have considered here is sufficient to capture the underlying physics of the DNA star tile self-assembly process. Unfortunately, it would be prohibitively expensive to perform such simulations using a brute-force all-atom approach, since both the time and the length scales involved are far too large. As a compromise, we use what is still a coarse-grained, but much more realistic model of DNA, oxDNA,\cite{Ouldridge2010, Snodin2015} to make further progress. Even though a number of features of DNA have already been coarse-grained within this model, studying the formation of quasicrystals and phase behaviour with oxDNA is still too computationally intractable to be feasible to attempt in full. However, what we can study is the behaviour of the basic quasicrystalline motifs that we observe in the patchy-particle simulations. We have confirmed from these simulations that the structures predicted by our `toy' patchy-particle model are reasonably well behaved and the patchy model does appear to capture the necessary fundamentals of the physical system. This is a very exciting result because it gives us a considerable degree of confidence that it might be possible in experiment to self-assemble a soft quasicrystal using DNA molecules. 

\section{Patchy-particle simulations}
\subsection{Model and methods}
Patchy-particle models have been used extensively to study a wide range of behaviours in computer simulations,\cite{Wilber2007, Noya2007,  Noya2010,  Williamson2011b, Doppelbauer2012b, Doye2007, VanDerLinden2012, Doppelbauer2010, Kern2003, Zhang2004, Bianchi2006, Sciortino2009, Romano2011b, Bianchi2011, Reinhardt2013, Whitelam2015c, Reinhardt2016b, Whitelam2016} including self-assembly and crystallisation, and represent one of the simplest types of `toy model' which can account for the complexity of behaviour seen in experiments on a number of colloidal systems.

\begin{figure}[t!]
\centering
\includegraphics{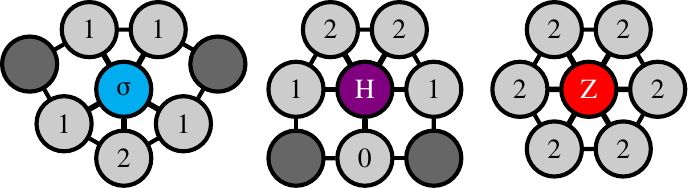}
 \caption{Neighbour classification of the $\upsigma$, H and Z environments. The nearest neighbours of the central particle are shown in light grey, and where applicable, the second-nearest neighbours in dark grey. The number given for each nearest neighbour specifies how many neighbours that particle shares with the central particle.}\label{fig-quasicrystal-neighClass}
\end{figure}

In our simulations, we use the Metropolis Monte Carlo scheme\cite{Metropolis1953} with volume moves\cite{Frenkel2002,Eppenga1984} and periodic boundary conditions. In simulations with multiple particle types, we furthermore allow moves in which two particles of distinct types are exchanged with one another in order to help facilitate equilibration.

We model particles with attractive patches using a simple angular modulation of the attractive part of the Lennard-Jones potential,
\begin{equation}
V(\mathbold{r}_{ij},\varphi_i,\varphi_j)=\begin{cases}
V^{\text{LJ}}(r_{ij}) & r_{ij}<\sigma_\text{LJ}, \\
V^{\text{LJ}}(r_{ij})  V^{\text{A}}(\hat{\mathbold{r}}_{ij},\varphi_i,\varphi_j) &  \sigma_\text{LJ}\le r_{ij} , \\
\end{cases}\label{patchy-potential}
\end{equation}
where $\mathbold{r}_{ij}$ is the interparticle vector connecting the particles $i$ and $j$, $r_{ij}$ is the magnitude of this vector, and $\varphi_i$ and $\varphi_j$ are the orientations of the particles $i$ and $j$, respectively. The Lennard-Jones potential is given by
\begin{equation}
V^{\text{LJ}}(r_{ij}) =
4 \varepsilon \left[\left(\frac{\sigma_\text{LJ}}{r_{ij}}\right)^{12}-\left(\frac{\sigma_\text{LJ}}{r_{ij}}\right)^{6}\right],\label{patchy-LJ}
\end{equation}
and we use a potential cutoff of $r_\text{cut}=3\sigma_\text{LJ}$ and shift the potential so that it equals zero at $r_\text{cut}$.  The angular modulation term in the potential is given by a product of gaussian functions,
\begin{equation}
V^{\text{A}}(\hat{\mathbold{r}}_{ij},\,\varphi_i,\,\varphi_j) = \max_{k,\,l}\mleft\{\exp\mleft[\frac{-\theta_{kij}^2}{2\,\sigma_\text{pw}^2}\mright]\exp\mleft[\frac{-\theta_{lji}^2}{2\,\sigma_\text{pw}^2}\mright]\mright\},\label{eqn-angmod}
\end{equation}
where $\sigma_\text{pw}$ is a parameter reflecting the patch `width' and $\theta_{kij}$ is the angle between the patch vector of patch $k$ on particle $i$ and the interparticle vector $\mathbold{r}_{ij}$. The product of gaussian functions is evaluated over all possible patch pairs $\{k,\,l\}$, and the optimum combination is chosen, i.e.~a pair of particles can only interact via that pair of patches which is most energetically favourable.

In simulations with multiple patch types, the angular modulation of Eqn~\eqref{eqn-angmod} is modified to include a prefactor that depends on the interaction matrix of the two patches $k$ and $l$ considered. In all simulations considered here, the matrix elements of this matrix are either zero or unity, i.e.~all patches that interact have the same strength, and patches that do not interact do not contribute at all to the energy in the attractive part of the Lennard-Jones potential. The most energetically favourable pair of patches is still chosen in the computation of the angular modulation in Eqn~\eqref{eqn-angmod}.

In order to characterise the structures we observe in our simulations, we classify each particle according to its nearest-neighbour environment.\cite{VanDerLinden2012, Reinhardt2013} To do this, we determine the neighbours of each particle, using a simple spherical cutoff of $1.38\,\sigma_\text{LJ}$,\cite{VanDerLinden2012} and then determine how many neighbours each neighbouring particle shares with the particle we are classifying. We classify particles into three distinct types of environment, $\upsigma$, H and Z, with common neighbour signatures of $\{21111\}$, $\{22110\}$ and $\{222222\}$ respectively, as illustrated in Fig.~\ref{fig-quasicrystal-neighClass}. This labelling corresponds to the equivalent Frank--Kasper phases.\cite{Frank1958, Frank1959}  In all simulation snapshots shown in the next section, particles are coloured according to their classification following the colour-coding shown in Fig.~\ref{fig-quasicrystal-neighClass}, namely cyan ($\upsigma$), violet (H) and red (Z), with particles whose environments give any common neighbour signature not listed above depicted in green. In simulations with multiple particle types, the base particle colour corresponding to Fig.~\ref{fig-quasicrystal-neighClass} is mixed with varying amounts of black for particles of types A and C (as defined below) in order to help to distinguish them.  

\subsection{Results and discussion}
\begin{figure}[b]
\centering
\includegraphics{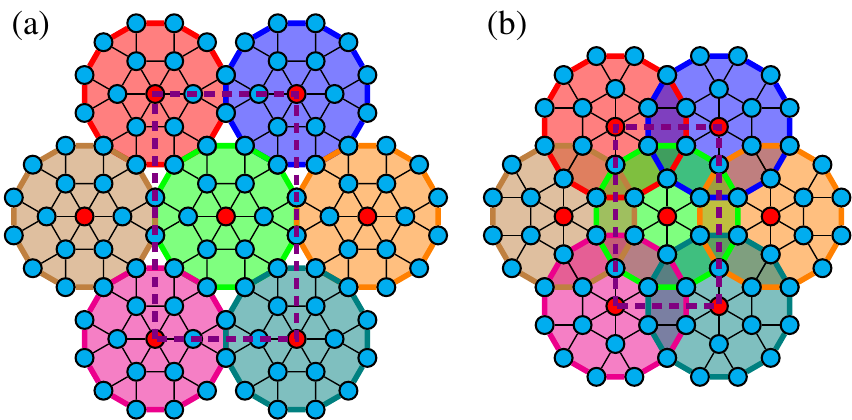}
 \caption{Two of the most common local structures in the dodecagonal quasicrystals studied previously,\cite{VanDerLinden2012, Reinhardt2013} with (a)~edge-sharing and (b)~overlapping dodecagonal motifs. For clarity, the individual dodecagonal motifs are shaded in distinct colours to emphasise their overlap. These individual motifs are rotated by \ang{90} in (b). Each particle is colour-coded based on its classification as per Fig.~\ref{fig-quasicrystal-neighClass}. A possible unit cell for each approximant crystal is outlined in violet.}\label{fig-quasicrystal-dodecagonal-unitcell}
\end{figure}

In all simulations reported here, the patch width was chosen to be $\sigma_\text{pw} = \SI{0.3}{\radian}$. For the patchy potential introduced in Eqn~\eqref{patchy-potential}, this is quite a narrow patch width,\cite{Doye2007} making interactions very angularly dependent. Such a narrow patch width allows us to account for the relatively highly directional nature of DNA multi-arm motif structures and prevent competition between environments of different valencies. However, for precisely the same reason, this choice of patch width leaves us in a region of parameter space where, for pentavalent particles we considered previously,\cite{VanDerLinden2012, Reinhardt2013} the quasicrystal was \emph{not} thermodynamically stable, as the patches are so narrow that it is not possible for six neighbours to bond competitively: instead, the $\upsigma$ phase was stable for pentavalent particles under these conditions.\cite{Reinhardt2013}

We note that in the quasicrystals we studied previously,\cite{VanDerLinden2012, Reinhardt2013} the most common structural feature was a series of edge-sharing dodecagonal motifs; one of the most common such motifs is shown in Fig.~\refSub{a}{fig-quasicrystal-dodecagonal-unitcell}.  In the `unit cell' of the approximant crystal corresponding to this motif, also illustrated in Fig.~\refSub{a}{fig-quasicrystal-dodecagonal-unitcell}, there are two particles in a hexagonal (Z) environment and 24 particles in a $\upsigma$ environment. Since the former correspond to particles having six neighbours and the latter to only five, we can surmise that in order to achieve our goal of assembling quasicrystals with particles with a narrow patch width, including both hexa- and pentavalent patchy particles in a simulation box in a ratio of $1 : 12$ would be a sensible choice.

\begin{figure}
\centering
\includegraphics{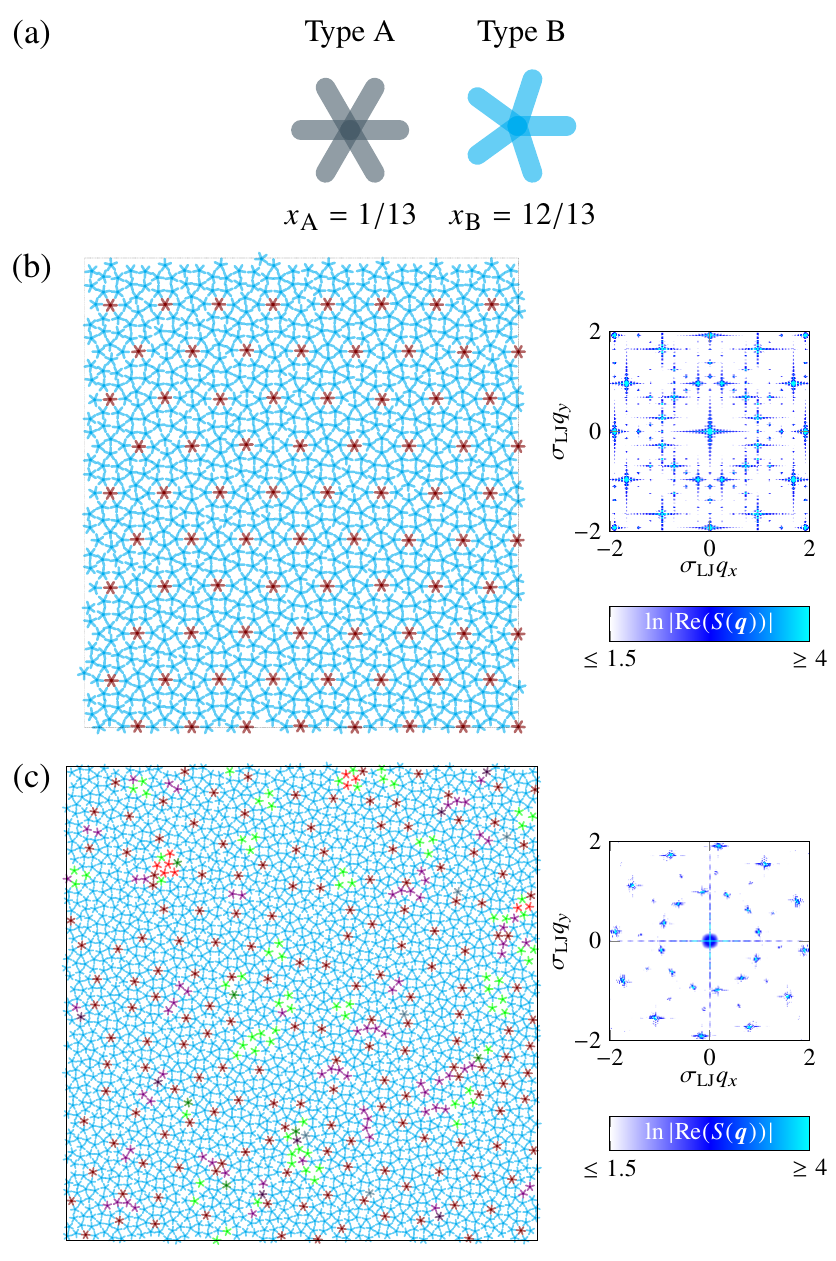}
 \caption{Non-specific patchy particles assembling into a quasicrystal. (a)~Particle types and the mole fraction of each type used in simulations. (b)~An equilibrated quasicrystalline approximant corresponding the structure of Fig.~\refSub{a}{fig-quasicrystal-dodecagonal-unitcell}. 1040 particles in total. $k_\text{B}T/\varepsilon=0.1$, $\sigma_\text{LJ}^2 \beta P = 1.5$. (c)~A dodecagonal quasicrystalline configuration that was obtained from a cooling run, starting from a liquid state, as the temperature was gradually decreased to $k_\text{B}T/\varepsilon=0.16$. $\sigma_\text{LJ}^2 \beta P = 1.5$. 2496 particles in total. In (b) and (c), the diffraction pattern computed for the configuration depicted is also shown. $S(\mathbold{q})$ is the structure factor evaluated at the reciprocal space vector $\mathbold{q}=(q_x,\,q_y)$.}\label{fig-quasicrystal-dodecagonal-nonspecific}
\end{figure}

\begin{figure}
\centering
\includegraphics{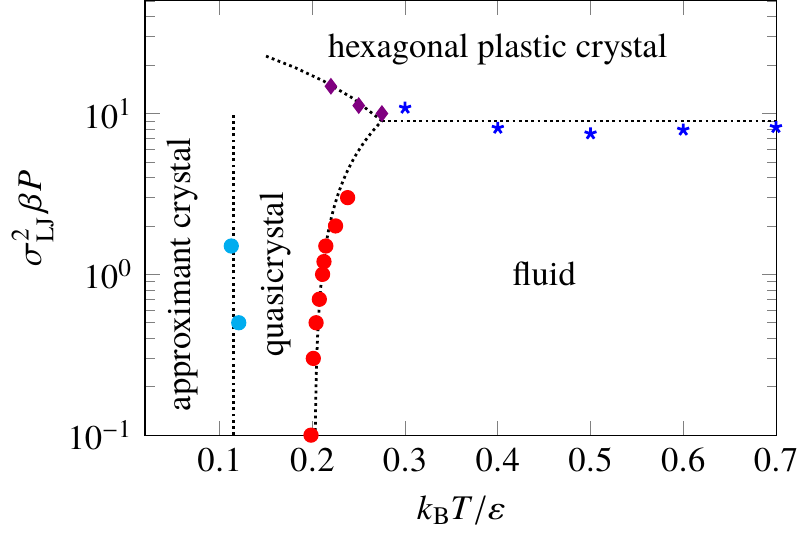}
 \caption{An approximate phase diagram for the non-specific quasicrystal-forming system. The approximant--quasicrystal coexistence data points come from free-energy calculations; the remaining points all come from direct-coexistence or brute-force simulations. The dotted lines are guides to the eye only. The boundary between the quasicrystal and the quasicrystalline approximant is likely to be an overestimate in temperature, and the true region of stability for the quasicrystal is likely to be somewhat larger than shown here.}\label{fig-binary-quasi-PW0.3-phasediagram}
\end{figure}

We have run simulations with a mixture of hexa- and pentavalent patches in this ratio [Fig.~\ref{fig-quasicrystal-dodecagonal-nonspecific}]. A quasicrystal phase forms spontaneously when the liquid phase is cooled [Fig.~\refSub{c}{fig-quasicrystal-dodecagonal-nonspecific}]; its quasicrystallinity is confirmed by the diffraction pattern shown in Fig.~\refSub{c}{fig-quasicrystal-dodecagonal-nonspecific}, which clearly exhibits dodecagonal symmetry: since this dodecagonal order is coherent through the whole of the simulation box, this is a single quasicrystal. There are several features of this quasicrystalline configuration worth noting. The large majority of hexavalent patches are at the centres of dodecagons, and many of these dodecagons are arranged locally in the motif of Fig.~\refSub{a}{fig-quasicrystal-dodecagonal-unitcell}. However, a common alternative motif for a quasicrystalline approximant is depicted in Fig.~\refSub{b}{fig-quasicrystal-dodecagonal-unitcell}, based on overlapping, rather than edge-sharing, dodecagonal motifs. We can see that there are sections of the structure in Fig.~\refSub{c}{fig-quasicrystal-dodecagonal-nonspecific} which are locally like both the edge-sharing and the overlapping approximants, both of which form a triangular lattice with longer and shorter distances between the `vertices' of the lattice. However, it is worth noting that there are also a number of other motifs of dodecagon centres, such as rectangular and isosceles triangular ones, which feature in Fig.~\refSub{c}{fig-quasicrystal-dodecagonal-nonspecific}.

As we decrease the temperature further still, the quasicrystalline approximant is expected to become the stable phase, as the configurational entropy of the quasicrystal becomes relatively less important compared to the enthalpic stability of the approximant.\cite{Reinhardt2013} In order to check that the quasicrystal is thermodynamically stable at intermediate temperatures, rather than simply a kinetic product, we have computed the free energies of the competing phases. To find the free energy of the edge-sharing approximant [Fig.~\refSub{b}{fig-quasicrystal-dodecagonal-nonspecific}], we used Frenkel--Ladd integration\cite{Frenkel1984} at $k_\text{B}T/\varepsilon=0.1$ and $\sigma_\text{LJ}^2 \beta P = 0.5$ (where $\beta=1/k_\text{B}T$), and, for consistency checking, also at $\sigma_\text{LJ}^2 \beta P = 1.5$, and then integrated this free energy along iso-$(\beta P)$ curves.\cite{Vega2008} We found the free energy of the fluid phase by integrating from the ideal gas, bearing in mind that it is a multicomponent gas. The free energy of the quasicrystal was then set by equating it to the free energy of the fluid at the point at which the quasicrystal and the fluid phase are at equilibrium; we determined this condition by direct-coexistence simulations.\cite{Reinhardt2013} By finding where the free energy curves of the approximant and the quasicrystal cross, we can obtain the relevant coexistence curve between the two phases. Finally, at very high pressures, the hexagonal (Z) plastic crystal phase dominates because its density is larger,\cite{Reinhardt2013} whether or not all neighbour--neighbour interactions can be satisfied. However, bearing in mind that the corresponding DNA systems are very dilute solutions, if we work at reasonable pressures, the Z phase does not need to be considered further.

At such reasonable pressures (i.e.~for $\sigma_\text{LJ}^2\beta P \lesssim 10$), the quasicrystal is thermodynamically stable relative to the fluid at temperatures below about $k_\text{B}T/\varepsilon = 0.2$. The approximant crystal takes over below about $k_\text{B}T/\varepsilon = 0.1$.  However, it is worth bearing in mind that it is considerably more difficult to equilibrate the quasicrystal at low temperatures when there is a mixture of hexa- and pentavalent components in the system than in the previously considered work, and the precise coexistence point between the approximant and the quasicrystal depends quite strongly on how well equilibrated the quasicrystal is: the lower temperature limit thus gives the minimum region of stability of the quasicrystal, and its region of stability is expected to be somewhat larger still in practice, since, unlike for pure pentavalent particles, the quasicrystal could largely be fully bonded, and so we expect little difference in energy between such a configuration and the approximant. An approximate phase diagram is shown in Fig.~\ref{fig-binary-quasi-PW0.3-phasediagram}.\footnote{It ought to be borne in mind that this phase diagram relates to the two-dimensional set-up considered in the patchy-particle simulations; the true phase diagram of DNA tile systems, which are at equilibrium with a three-dimensional solution of building blocks, is unlikely to feature a two-dimensional fluid, and so the phase diagram of Fig.~\ref{fig-binary-quasi-PW0.3-phasediagram} is only intended to provide a rough idea of what the actual phase behaviour might be.} Interestingly, the quasicrystalline phase is stable over a wider range of temperatures than for the pure pentavalent patchy particle system we previously studied;\cite{Reinhardt2013} this is because the energy difference between the quasicrystal and the approximant in the current system is smaller than that between the quasicrystal and the $\upsigma$ phase for the pentavalent particles.

Unlike in the single-component phase diagram considered in Ref.~\citenum{Reinhardt2013}, the low-temperature phase, at which the configurational entropy afforded by the quasicrystal is no longer as important as it is to maximise the bonding, is the approximant crystal rather than the $\upsigma$ phase, since the presence of hexavalent particles means that bonding cannot be maximised in the $\upsigma$ geometry for all particles. However, it would alternatively be possible that the mixture may phase separate into a $\upsigma$ and a Z phase. We have confirmed that the zero-temperature enthalpy of the approximant phase is lower than the enthalpy of a $1 : 12$ mixture of Z and $\upsigma$ phases comprising solely hexa- and pentavalent particles respectively at all pressures considered, even with no interfacial energy penalty imposed. The approximant crystal is therefore stable with respect to phase separation at zero temperature.

\begin{figure}
\centering
\includegraphics{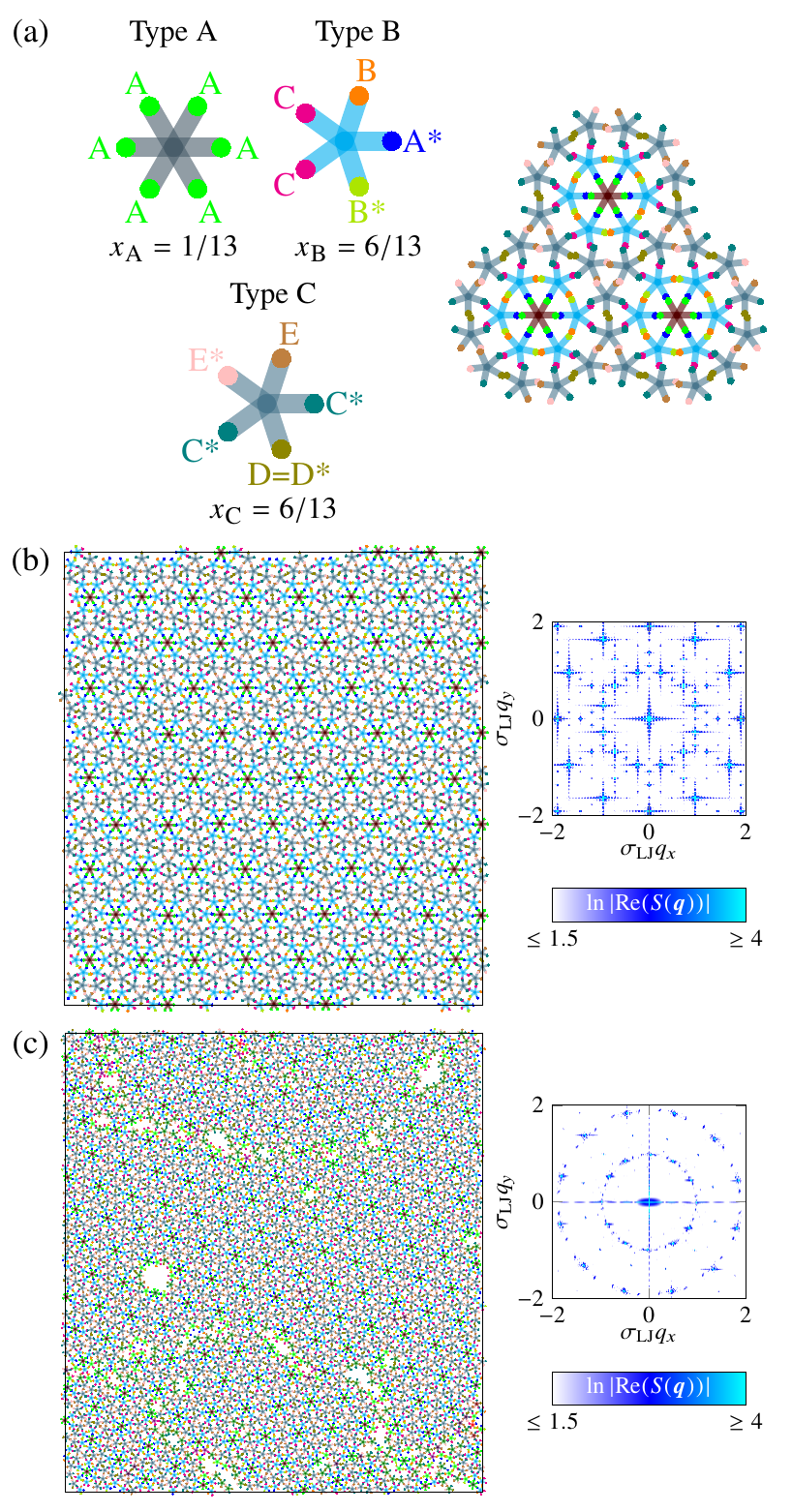}
 \caption{Fully specific patchy particles designed to form the quasicrystalline approximant of Fig.~\refSub{a}{fig-quasicrystal-dodecagonal-unitcell}. (a) Particle types and the mole fraction of each type used in simulations. Patches only interact with complementary patches, indicated by an asterisk. The basic motif of edge-sharing dodecagons with explicit patch--patch interactions is also shown. (b) An equilibrated approximant crystal corresponding to a structure in which all bonding interactions are satisfied [Fig.~\refSub{a}{fig-quasicrystal-dodecagonal-unitcell}]. 1040 particles in total. $k_\text{B}T/\varepsilon=0.1$, $\sigma_\text{LJ}^2 \beta P = 1.5$. (c) A crystalline configuration obtained from a cooling run, starting from the fluid, in which the temperature was gradually decreased to $k_\text{B}T/\varepsilon=0.16$. $\sigma_\text{LJ}^2 \beta P = 1.5$. 2496 particles in total. For (b) and (c), the corresponding diffraction patterns are also shown.}\label{fig-quasicrystal-fully-specific-edgesharing}
\end{figure}

The stability of the quasicrystal over a range of conditions demonstrates that it is possible to set up a competition between penta- and hexavalent co-ordination by means other than having a large patch width. In particular, in previous work, the quasicrystal was never found to be stable for patch widths below $\sigma_\text{pw}\approx 0.45$.\cite{Reinhardt2013} By contrast, we have shown here that a quasicrystal phase is thermodynamically stable even if the patch width is considerably narrower (i.e.~$\sigma_\text{pw}=0.3$). This is very encouraging if our aim is to construct quasicrystals using DNA multi-arm motifs. However, at this stage, the set-up we have considered completely ignores one of the most important reasons why DNA is so popular in experiment: using DNA makes it very easy to design mutually orthogonal interactions. Indeed, the multi-arm DNA tiles equivalent to the patchy particles that we have considered thus far, where all the sticky ends at the end of the arms have a favourable interaction with all other sticky ends, might be rather more difficult to self-assemble than envisaged because growth might be arrested as multiple undesigned bonds are formed, and so it is prudent to investigate whether it remains possible to form quasicrystals if the patches are made to be more specific. However, while the kinetics of self-assembly may be less frustrated as the interactions are made more specific, it is important to bear in mind that it is in the nature of quasicrystals that they are not completely ordered (indeed, this is what affords them their additional entropic stability): it is not possible, by construction, to have a set of interactions for which the fully bonded configuration is uniquely determined to be the quasicrystal.

\begin{figure}
\centering
\includegraphics{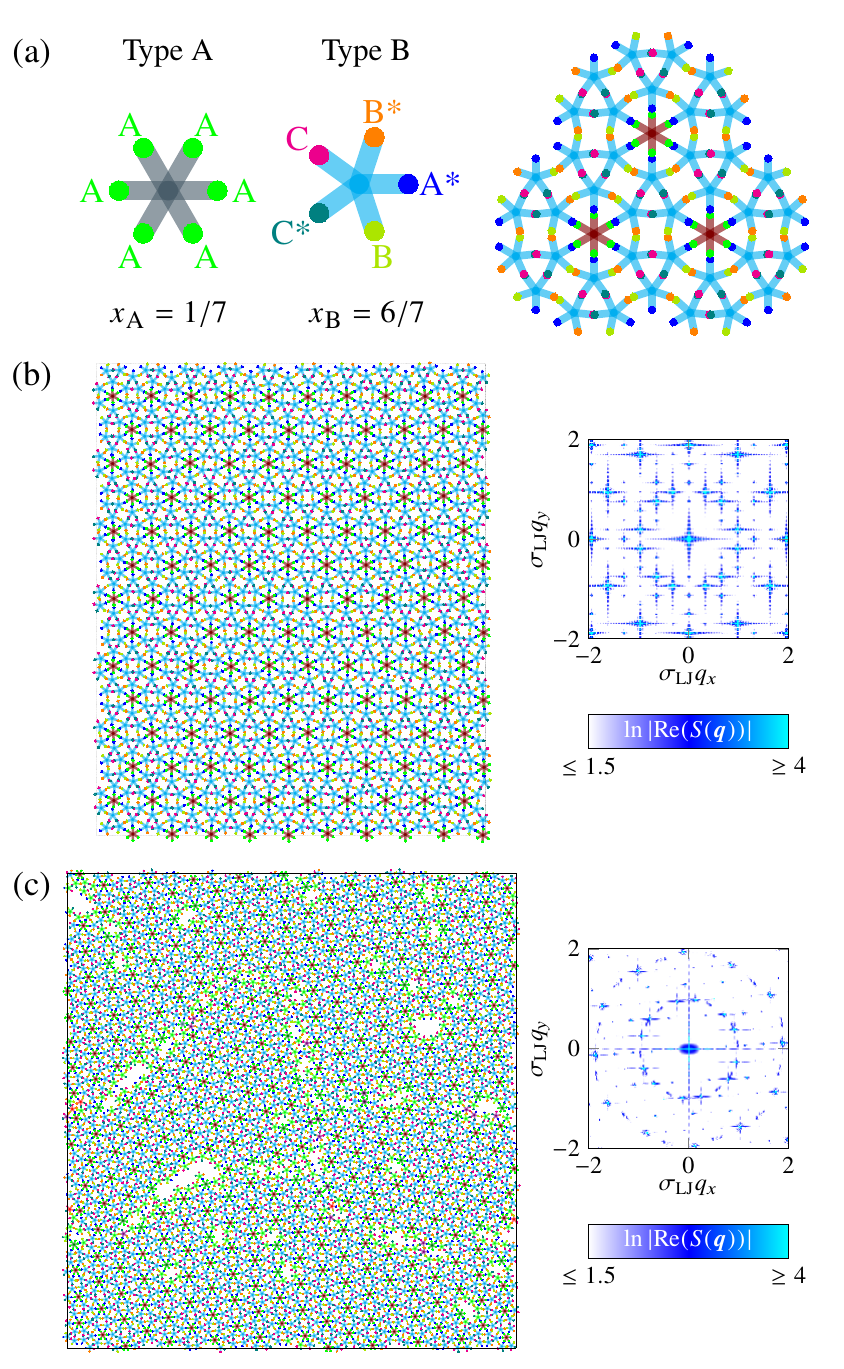}
 \caption{Fully specific patchy particles designed to form the quasicrystalline approximant of Fig.~\refSub{b}{fig-quasicrystal-dodecagonal-unitcell}. (a) Particle types and the mole fraction of each type used in simulations. Patches only interact with complementary patches, indicated by an asterisk. The basic motif of overlapping dodecagons with explicit patch--patch interactions is also shown. (b) An equilibrated approximant crystal corresponding to a structure in which all bonding interactions are satisfied [Fig.~\refSub{b}{fig-quasicrystal-dodecagonal-unitcell}]. 980 particles in total. $k_\text{B}T/\varepsilon=0.16$, $\sigma_\text{LJ}^2 \beta P = 1.5$. (c) A crystalline configuration obtained from a cooling run, starting from the fluid, in which the temperature was gradually decreased to $k_\text{B}T/\varepsilon=0.16$. $\sigma_\text{LJ}^2 \beta P = 1.5$. 2492 particles in total. For (b) and (c), the corresponding diffraction patterns are also shown.}\label{fig-quasicrystal-fully-specific-overlapping}
\end{figure}

As a first step in exploring the factors that govern quasicrystal self-assembly in binary mixtures corresponding to DNA star tiles, we consider the interaction set required to form two of the quasicrystalline approximants considered above [Fig.~\ref{fig-quasicrystal-dodecagonal-unitcell}], before considering how these interactions could be relaxed to allow the variety of environments typical of our target dodecagonal quasicrystal to form. Let us first consider how we can make every distinct type of interaction that can be identified in the unit cell of Fig.~\refSub{a}{fig-quasicrystal-dodecagonal-unitcell} different. This can be achieved by making the pentavalent particles of two different types, as depicted in Fig.~\refSub{a}{fig-quasicrystal-fully-specific-edgesharing}. In this set-up, patches only interact with complementary patches denoted by the same letter and an asterisk; other patch pairs do not interact at all.  The unit cell of the edge-sharing approximant of  Fig.~\refSub{a}{fig-quasicrystal-dodecagonal-unitcell} can readily be identified in the approximant shown in Fig.~\refSub{b}{fig-quasicrystal-fully-specific-edgesharing}. However, whilst the approximant is stable in roughly the same conditions as it was before, the quasicrystal is not expected to form with such specific interactions. Even when cooled very slowly, kinetic products such as that shown in Fig.~\refSub{c}{fig-quasicrystal-fully-specific-edgesharing} are obtained. Whilst the underlying approximant crystal ordering can certainly be identified in this figure, the fact that these dodecagonal motifs are not orientated in the same way throughout the simulation box means that large gaps must be left in order to reduce the strain in the system, and since the bonding is so specific, no particles can be used to `glue' the different regions together. The resulting structure is therefore, unsurprisingly, full of defects: it is not a single crystal, but has several crystalline domains with grain boundaries between them. The fact that there are multiple crystallites is confirmed by the smeared out diffraction pattern that involves a superposition of patterns from the different crystallites [Fig.~\refSub{c}{fig-quasicrystal-fully-specific-edgesharing}].

The alternative approximant motif of Fig.~\refSub{b}{fig-quasicrystal-dodecagonal-unitcell} is based on overlapping rather than edge-sharing dodecagonal motifs. In the  `unit cell' of this alternative approximant crystal, there are 2 particles in a hexagonal (Z) environment and 12 particles in a $\upsigma$ environment, necessitating a ratio of $1 : 6$ of hexa- and pentavalent patchy particles in the simulation box. If the bonding interactions are made fully specific with this alternative set-up, as shown in Fig.~\refSub{a}{fig-quasicrystal-fully-specific-overlapping}, we can again stabilise the approximant crystal [Fig.~\refSub{b}{fig-quasicrystal-fully-specific-overlapping}]. However, similar to the edge-sharing fully specific system, on cooling, multiple nucleation events occur, leading to multiple crystallites of the overlapping approximant separated by highly defective grain boundaries.

\begin{figure}
\centering
\includegraphics{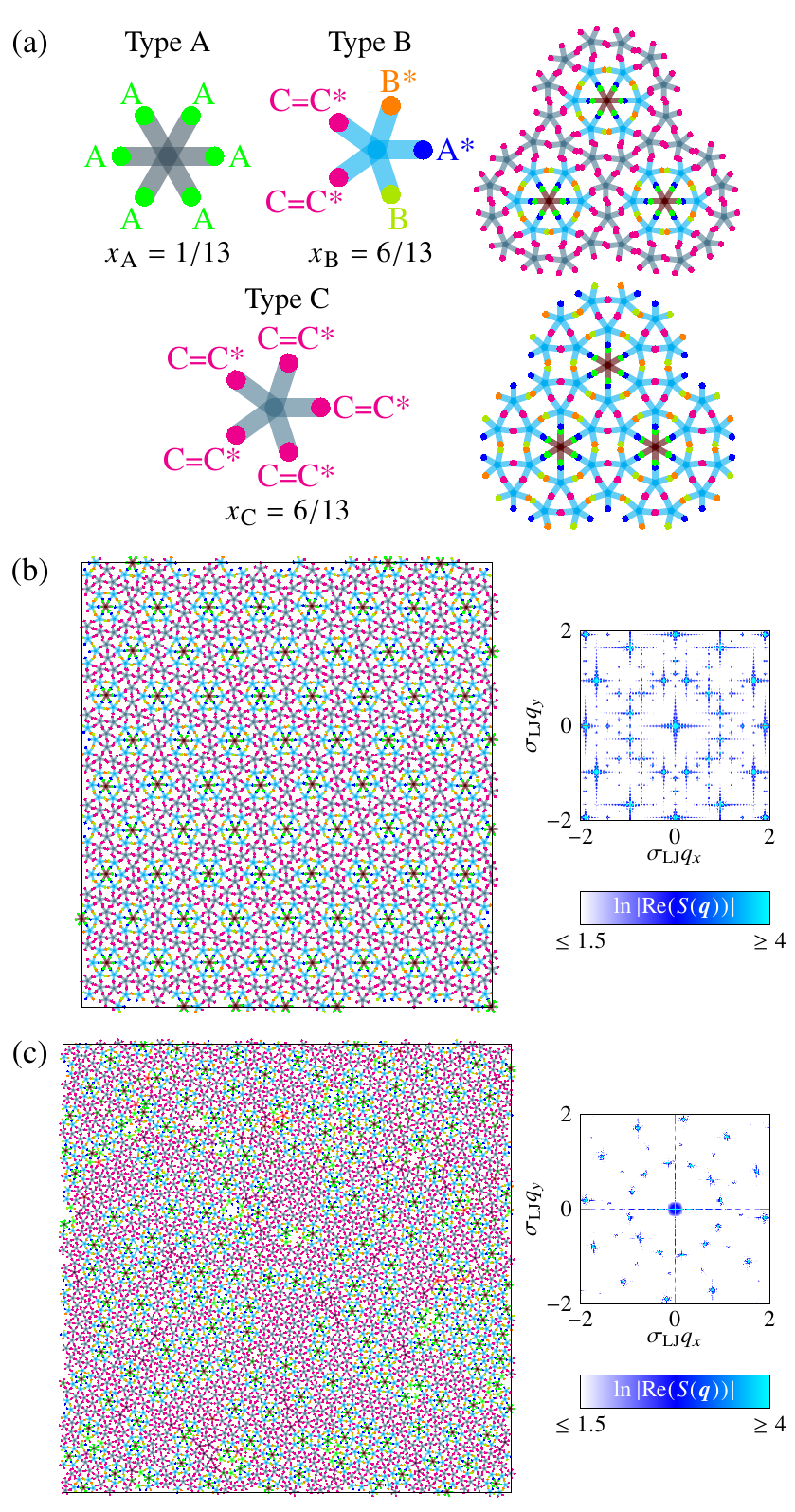}
 \caption{Patchy particles of `intermediate' specificity that are designed to allow quasicrystal formation. (a) Particle types and the mole fraction of each type used in simulations. Patches only interact with complementary patches, indicated by an asterisk. The basic motifs of both edge-sharing and overlapping dodecagons with explicit patch--patch interactions are also shown. (b) An equilibrated edge-sharing approximant crystal. 1040 particles in total. $k_\text{B}T/\varepsilon=0.1$, $\sigma_\text{LJ}^2 \beta P = 1.5$. (c) A quasicrystal resulting from a cooling run in which the temperature was gradually decreased to $k_\text{B}T/\varepsilon=0.16$. $\sigma_\text{LJ}^2 \beta P = 1.5$. 2496 particles in total. For (b) and (c), the corresponding diffraction patterns are also shown.}\label{fig-quasicrystal-intermed-specific}
\end{figure}

In order to improve the kinetics whilst retaining enough plasticity in the interactions to allow quasicrystals, rather than just crystals, to form, we can aim to strike a balance between the full specificity considered in simulations illustrated by Figs~\ref{fig-quasicrystal-fully-specific-edgesharing} and \ref{fig-quasicrystal-fully-specific-overlapping} on the one hand and the completely non-specific bonding of Fig.~\ref{fig-quasicrystal-dodecagonal-nonspecific}. To this end, we have considered an alternative set-up in which we allow a competition between the overlapping and edge-sharing dodecagonal motifs to be set up. As illustrated in Fig.~\refSub{a}{fig-quasicrystal-intermed-specific}, by permitting the outlying `type C' particles to bond freely with two of the patches of `type B' particles, we can assemble either structures analogous to those of Fig.~\ref{fig-quasicrystal-fully-specific-edgesharing}, involving all three types of particle, or of Fig.~\ref{fig-quasicrystal-fully-specific-overlapping}, involving only particles of types A and B, with the excess of particles of type C forming regions of $\upsigma$-environments that can fill the gaps between `approximant' motifs that are orientated in different ways. We have chosen the composition of particle types in this set-up to be the same as the one we considered for the edge-sharing approximant system of Fig~\ref{fig-quasicrystal-fully-specific-edgesharing}.

With this set-up, cooling a liquid again results in a quasicrystalline phase, such as that shown in Fig.~\refSub{c}{fig-quasicrystal-intermed-specific}, with the corresponding diffraction pattern confirming its dodecagonal symmetry. As in the non-specific case of Fig.~\ref{fig-quasicrystal-dodecagonal-nonspecific}, there is again a variety of ways in which the dodecagons pack in addition to the two triangular lattice patterns of the approximants of Fig.~\ref{fig-quasicrystal-dodecagonal-unitcell}.  To verify that the quasicrystal is thermodynamically stable, we have repeated the calculations considered above for the non-specific case at $\sigma_\text{LJ}^2 \beta P = 1.5$. In particular, we have computed the free energies of the edge-sharing and overlapping approximants as well as the $\upsigma$ phase using Frenkel--Ladd integration. The free energy of the edge-sharing approximant matches the free energy of a system combining 7 parts of the overlapping approximant and 6 parts of the $\upsigma$ phase; this ratio accounts for the excess of `type C' particles when an overlapping approximant is formed at the considered composition of particles of types A, B and C [cf.~Fig.~\refSub{a}{fig-quasicrystal-intermed-specific}]. At temperatures above $k_\text{B}T/\varepsilon \approx 0.1$, the quasicrystal's free energy is lower than that of the approximants, confirming that it is the thermodynamically stable phase across a range of temperatures.

The fact that the quasicrystalline phase is thermodynamically stable for this system of `intermediate' specificity suggests that a DNA star tile system with interactions chosen in this way would be the most likely to result in a two-dimensional soft DNA-based quasicrystal. However, it is certainly the case that we have ignored a number of considerations when abstracting the system to the toy-model level considered here. For example, it is not at all clear \textit{a priori} that dodecagonal motifs comprising DNA star tile would be sufficiently planar to permit the growth of suitably large two-dimensional quasicrystalline structures, although this consideration may be mitigated to a large extent by performing the self-assembly on a surface. In order to address this point, we turn briefly to a more realistic potential of DNA molecules themselves.

\section{Simulating DNA tile arrays with a realistic model}
\subsection{OxDNA model and methods}

OxDNA\cite{Ouldridge2011,Snodin2015} is a coarse-grained DNA model at the nucleotide level that allows the simulation of systems of large numbers of nucleotides.  The nucleotides are modelled as rigid bodies interacting with a series of effective interactions (the solvent is not explicitly modelled) that account for hydrogen bonding between Watson--Crick base pairs, stacking between bases, electrostatic repulsion between the phosphates, excluded volume and chain connectivity. These interactions have been fitted to reproduce the thermodynamics of hybridisation and the structural and mechanical properties of both single-stranded and double-stranded DNA. Here, we use the second version of the model (`oxDNA2') that includes fine-tuned properties to reproduce better the properties of large nanostructures -- in particular DNA origamis.\cite{Snodin2015}

\begin{figure*}
\includegraphics{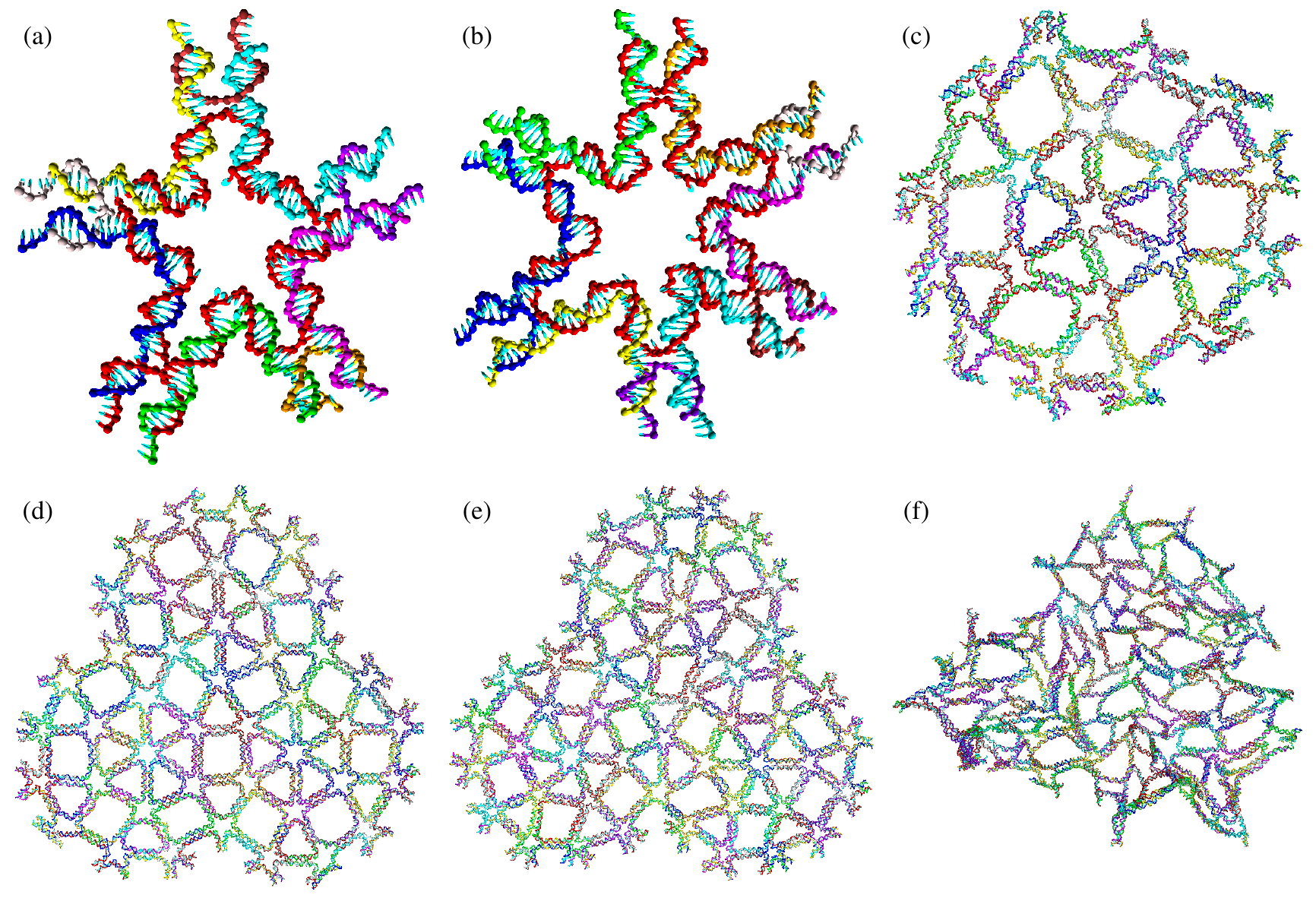}
\caption{OxDNA representations of (a) a five-arm tile, (b) a six-arm tile, (c), (d), (e) quasicrystal-like motifs of increasing size when adsorbed onto a surface and (f) the largest motif when in solution. In (a) and (b), flat configurations have deliberately been chosen to most clearly illustrate the design and topology of the star tiles. The simulations of these structures were all performed at \SI{22}{\celsius} and a salt concentration of \SI{0.5}{\mole\per\cubic\deci\metre}.}\label{fig-oxDNA-star-tiles}
\end{figure*}

The oxDNA model is the most widely used coarse-grained model of DNA at the nucleotide level, and has been used to study the biophysical properties of DNA, a wide variety of DNA nanotechnology systems and applications in soft matter materials.\cite{Doye2013} These applications have confirmed the model's robustness and quantitative accuracy (e.g.~Ref.\ \onlinecite{Srinivas2013}'s reproduction of six orders of magnitude variation in the kinetics of strand displacement). Particularly relevant to the current application is the model's ability to account for the structural properties of both bulged duplexes,\cite{Schreck2015} a motif that is crucial to the properties of DNA star tiles, and polyhedral assemblies of star tiles,\cite{Schreck2016} and to rationalise how the kinetics of star tile self-assembly can be controlled through the size of the bulges they contain.\cite{Schreck2016b}

Our aim here is to use oxDNA to explore the structural stability of quasicrystalline arrays made out of DNA star tiles. To achieve this, we first need to generate starting initial structures for these arrays. We do this by first designing the arrays using vhelix,\cite{vhelix} a recent DNA nanostructure design programme that allows free placement of the component DNA helices in space rather than on a lattice.\cite{Benson2015} We then convert the vhelix design into a starting geometry for the oxDNA model. This geometry is not yet a suitable starting point for molecular dynamics simulations, as it may have particle overlaps or extended bonds that give rise to unreasonably large energies and very large forces. We therefore first relax the structure using a steepest-descent-like minimisation technique.  The details of these procedures will be described elsewhere.\cite{Romano2016}

The structures are then simulated with a molecular dynamics algorithm employing an Andersen-like thermostat\cite{Russo2009} both to keep the temperature constant and to generate diffusive motion of the nucleotides, as is appropriate for molecules in solution.  As the systems we study contain thousands of nucleotides (the largest has 50 tiles and \num{22704} nucleotides), to make the simulations feasible on a reasonable time scale, they are run on GPUs using a specially developed code.\cite{Rovigatti2015} We consider systems of tiles both when free in solution, as is typical during the assembly process, and when adsorbed on a surface, as is the case when visualised by some type of microscopy (e.g.~AFM). The interaction of the nucleotides with the surface is modelled with a simple one-dimensional Lennard-Jones interaction that depends only on the distance of a nucleotide from the surface.

\subsection{Results and discussion}

Our aim here is use oxDNA to check whether there might be any structural reasons why the dodecagonal quasicrystals that we have seen for the above patchy particles might not be realisable using DNA star tiles. Previous experiments do not suggest any obvious hindrances. For example, both five-arm and six-arm tiles have been produced and found to assemble into two-dimensional crystalline arrays, forming $\upsigma$\cite{Zhang2008} and hexagonal\cite{He2006} crystals, respectively. Furthermore, mixtures of three- and four-arm tiles with specifically designed interactions have been shown to be able to produce more complex crystal structures.\cite{Zhang2013,Zhang2016} One possible complication is that, if the tiles are not flat, but rather the arms possess an intrinsic preference to bend in a given direction, then if all tiles face in the same direction, this has the potential to lead to curvature in the resulting structure that could hinder the assembly of an extended two-dimensional structure. One solution is to design the tiles so that they alternate in orientation, and any curvature cancels out,\cite{He2005b} but this approach is not available for structures that possess polygons with an odd number of edges. However, in the examples above,\cite{He2006,Zhang2008,Zhang2013,Zhang2016} conditions and designs were still found for which the assembly of extended two-dimensional arrays dominated over the formation of finite closed objects (e.g.~icosahedra for five-arm star tiles\cite{Zhang2008}). Furthermore, self-assembly of DNA star tiles on a surface has also been shown to be possible.\cite{Sun2009}

Example five- and six-arm star tiles are illustrated in Figs~\refSub{a}{fig-oxDNA-star-tiles} and \refSub{b}{fig-oxDNA-star-tiles}. The tiles consist of a long central strand, five or six medium-length strands that bridge two arms and five or six short strands which bind at the ends of each arm. The bulges on the long strand between the arms provide flexibility, allowing the arms to bend back on themselves. In the current examples, there are four nucleotides in the bulges, the same as was used experimentally to produce extended structures with five- and six-arm tiles.\cite{He2006,Zhang2008}

We have constructed DNA analogues of three example motifs that are important for the quasicrystalline structures observed in our patchy-particle simulations, namely a dodecagon, three overlapping dodecagons, and three edge-sharing dodecagons. The simulations of these structures showed that they are all stable at room temperature with the correct topology of the network maintained throughout the simulation. Example configurations that have been adsorbed on a surface [Figs~\refSub{c}{fig-oxDNA-star-tiles}--\hyperref[fig-oxDNA-star-tiles]{(e)}] clearly show the expected structures. Due to the flexibility of the star tiles, the quadrilaterals in the network need not be perfectly square. Furthermore, bending is not always localised to the bulge regions of the tiles, and can sometimes occur at the four-way junctions in the arms leading to further distortions from the idealised geometries, even for the triangles.

By contrast, when free in solution, although the topology of the network is retained, the motifs are highly fluxional and no longer look anything like the idealised two-dimensional target structure [Fig.~\refSub{f}{fig-oxDNA-star-tiles}]. This is both because of the inherent flexibility of the star tiles and because all tiles face in the same direction, so any tendencies for the arms to bend away from the plane in a preferred direction are additive. The non-planarity is probably also exacerbated by the relative small size of the motifs, as consequently a large number of the arms (over \SI{18}{\percent} even for the largest motif) are on the edge of the motif and lack the constraint of being connected to another tile. 

We should emphasise that the flexibility of the structures and the large fluctuations away from planarity do not affect the stability of the networks, nor do they mean that further self-assembly of the networks is necessarily hindered. When a new tile binds to a free arm on the edge of the motif, further binding is probably still most likely to occur in the intended way. However, the non-planarity may also make allowed, but not intended, arm binding more likely than when in a planar geometry because the relevant arms have been brought closer together by the non-planar fluctuations. This again emphasises the importance of annealing to facilitate the melting away of incorrect bindings.

\section{Conclusions}

We have performed simulations of patchy-particle systems with a narrow patch width to investigate the phase behaviour of particles that can be considered to be a `toy model' for DNA star tiles or analogous systems. We have confirmed our hypothesis, originally proposed in Ref.~\onlinecite{Reinhardt2013}, that mixtures of penta- and hexavalent particles can mutually associate to form stable dodecagonal quasicrystals. 

We have explored two designs which lead to quasicrystal formation. The first one of these involved no specificity in the patch--patch interactions: every patch could interact with every other patch in the system. In patchy-particle simulations using this set-up, the quasicrystalline phase formed readily and was shown to be thermodynamically stable using free-energy calculations. However, one might imagine that the self-assembly of the quasicrystal in a DNA context might be trickier with such non-specific interactions, since particles cannot `detect' whether they are in the correct environment from their initial interactions. For example, a hexavalent particle should ultimately end up at the centre of a dodecagonal motif, but has no way of ensuring that this will be the case when it is bonded to only a few of its neighbours. In order to ensure that the thermodynamically stable phase can form at experimentally accessible time scales, particles must be allowed to bind and unbind very readily: the self-assembly process must therefore take place at temperatures at which the driving force to form the target phase is very small. Since non-optimal configurations are less stable, a high temperature means that such motifs are likely to melt off, allowing the stable phase to form over time.

The kinetics of the toy patchy-particle systems are very fast and a quasicrystal forms readily in such a set-up. However, while our simulations certainly do not preclude the possibility of self-assembly being feasible in an equivalent DNA system, we can make use of the information content of the DNA to be more selective in the interparticle interactions and thus attempt to reduce the likelihood of kinetic traps precluding successful self-assembly in DNA tile systems. We must however remember that the stability of quasicrystals is largely down to their configurational entropy.\cite{Reinhardt2013, Oxborrow1993} Unlike for the increasingly complex DNA-based crystalline motifs that have been considered in the literature\cite{Zhang2015, Zhang2016} and as exemplified by the quasicrystalline approximants considered here, it is not therefore possible to design a set of interactions for which the fully bound ground-state configuration is uniquely specified to be a quasicrystal. Instead, we must design a set of interactions that provides sufficient freedom that will allow the full variety of motifs that are typical of the quasicrystal to form. In our design, specific interactions ensure that hexavalent particles are at the centre of dodecagons, but do not prescribe how these dodecagons associate. We have shown that quasicrystals are thermodynamically stable with a design of this kind, while the additional specificity of interactions should permit such structures to be more kinetically accessible than the equivalent non-specific system. An additional advantage of using more specific interactions is that they would counter against possible phase separation into penta- and hexavalent regions that has been seen for non-specific mixtures of three- and four-arm motifs.\cite{He2007} The specific interactions are also likely to reduce the competition with alternative closed polyhedral objects, but choosing the bulge size appropriately to inhibit these assembly pathways further is also likely to be important. 
We hope that the designs presented here might help to guide experimentalists in producing a DNA quasicrystal.

It may also be possible to design dodecagonal DNA crystals in other ways than the ones we have considered here. For example, we may consider the `duals' of the motifs considered above, where we interchange the nature of the vertices and faces in the structure. Since there is a bijective mapping between a structure and its dual,\cite{Frank1958} the dual of the quasicrystal we considered is also a quasicrystal. The resulting networks would only have tri- and tetravalent vertices. Although it is possible to design particles that would form the equivalents of the two approximant crystals that we consider here, it is rather less clear how to design specific but not overly constraining interactions which would encourage quasicrystal formation with such a set-up.

Finally, it is worth bearing in mind that there are several significant differences between the patchy particles we have considered here and real DNA multi-arm motifs. Firstly, DNA molecules are flexible, but have a fixed valency, whilst patchy particles have a fixed geometry and gain their flexibility in bonding through a patch width. We have made the patch width fairly narrow to ensure that the valency condition is maintained, but the dynamics of self-assembly may change if the patches were to be more flexible. Secondly, there is comparatively little excluded volume in DNA structures, whilst our patchy particles have a Lennard-Jones-style excluded volume interaction. However, since the structures of interest are stabilised by attractive rather than repulsive forces, we do not think that excluded volume effects would be particularly important. Thirdly, DNA assemblies can behave in rather more complex ways than we have considered here, since they exhibit a kind of structural co-operativity: curvature emerges from the assembly process, rather than being a property of isolated tiles.\cite{Schreck2016} However, several strategies exist for surface-mediated self-assembly,\cite{Hamada2009, Sun2009, Suzuki2015} and surface assembly may help to alleviate such problems with curvature.  Nevertheless, even if DNA tiles were to be assembled on a surface, rather than simply being adsorbed on a surface at the end of the process to visualise the structure, there is an important difference between a true two-dimensional system that we considered and assembly from a three-dimensional dilute solution on a surface. Finally, we used particle swap moves to help hexa- and pentavalent particles to find their preferred environment in patchy-particle simulations. Such Monte Carlo moves have no equivalent in real dynamics. However, in this case, the three-dimensional nature of the DNA assembly process may actually be beneficial, as in real DNA systems, even when adsorbed on a surface, the self-assembly is likely to be from a dilute three-dimensional solution, rather than from a two-dimensional fluid: we can envisage that the swap moves might correspond to adsorption and desorption of appropriate molecules from solution.

Of course, it may be possible to alleviate some of the concerns we have listed above by considering a more sophisticated model. For example, in order to address the issue with flexibility in the patchy-particle model, we could allow the patches to shift positions slightly in time, whilst maintaining a fixed valency by keeping the patches narrow. Alternatively, a more rigid DNA structure based on an origami approach\cite{Wang2016} might make it easier to design planar tiles in experiment. However, while there are certainly a number of simplifications and omissions in our patchy-particle approach, broadly speaking, as confirmed by the oxDNA simulations, the simple model we have considered appears to capture a sufficient amount of the underlying physics to serve as a good guide to the self-assembly behaviour of analogous DNA star tiles. We hope that our results will help to invigorate experimental efforts to produce a soft DNA-based quasicrystal.

\begin{acknowledgments}
This work was supported by the Engineering and Physical Sciences Research Council [Grants EP/I001352/1, EP/J019445/1]. We acknowledge the use of the University of Oxford Advanced Research Computing (ARC) facility in carrying out some of this work [\href{https://dx.doi.org/10.5281/zenodo.22558}{doi:10.5281/zenodo.22558}]. Supporting data are available at [TBD].
\end{acknowledgments}

\section*{References}

\vspace{-\baselineskip}

\end{document}